\documentclass[%
 aip,
 amsmath,amssymb,
 reprint,%
]{revtex4-1}

\usepackage{graphicx}
\usepackage{dcolumn}
\usepackage{bm}

\usepackage[utf8]{inputenc}
\usepackage[T1]{fontenc}
\usepackage{mathptmx}

\usepackage{booktabs} 
\usepackage{makecell}

\begin{document}

\preprint{AIP/123-QED}

\title[MPNNs for high-throughput polymer screening]{Message-passing neural networks for high-throughput polymer screening}

\author{Peter C. St.~John}
\affiliation{Biosciences Center, National Renewable Energy Lab, Golden, Colorado, USA}%
\email{peter.stjohn@nrel.gov}
\author{Caleb Phillips}
\author{Travis W. Kemper}
\affiliation{Computational Sciences Center, National Renewable Energy Lab, Golden, Colorado, USA}%
\author{A. Nolan Wilson}
\affiliation{National Bioenergy Center, National Renewable Energy Lab, Golden, Colorado, USA}%
\author{Yanfei Guan}
\affiliation{Department of Chemistry, Colorado State University, Fort Collins, Colorado, USA}%
\author{Michael F. Crowley}
\affiliation{Biosciences Center, National Renewable Energy Lab, Golden, Colorado, USA}%
\author{Mark R. Nimlos}
\affiliation{National Bioenergy Center, National Renewable Energy Lab, Golden, Colorado, USA}%
\author{Ross E. Larsen}
\affiliation{Computational Sciences Center, National Renewable Energy Lab, Golden, Colorado, USA}%

\begin{abstract}
Machine learning methods have shown promise in predicting molecular properties, and given sufficient training data machine learning approaches can enable rapid high-throughput virtual screening of large libraries of compounds.
Graph-based neural network architectures have emerged in recent years as the most successful approach for predictions based on molecular structure, and have consistently achieved the best performance on benchmark quantum chemical datasets.
However, these models have typically required optimized 3D structural information for the molecule to achieve the highest accuracy.
These 3D geometries are costly to compute for high levels of theory, limiting the applicability and practicality of machine learning methods in high-throughput screening applications.
In this study, we present a new database of candidate molecules for organic photovoltaic applications, comprising approximately 91,000 unique chemical structures.
Compared to existing datasets, this dataset contains substantially larger molecules (up to 200 atoms) as well as extrapolated properties for long polymer chains.
We show that message-passing neural networks trained with and without 3D structural information for these molecules achieve similar accuracy, comparable to state-of-the-art methods on existing benchmark datasets.
These results therefore emphasize that for larger molecules with practical applications, near-optimal prediction results can be obtained without using optimized 3D geometry as an input. 
We further show that learned molecular representations can be leveraged to reduce the training data required to transfer predictions to a new DFT functional.
\end{abstract}

\maketitle

\section{Introduction}

High-throughput computational screening offers the ability to explore large regions of chemical space for particular functionality, greatly enhancing the efficiency of material development efforts \citep{Jain_2013, Lopez_2016, Huang_2017}.
Due to its favorable balance between computational cost and chemical accuracy, density functional theory (DFT) has served as the workhorse of high-throughput computational material design.
However, while DFT sacrifices chemical accuracy for numerical efficiency, DFT calculations are still too slow to screen the vast combinatorial landscape of potential chemical structures \citep{Ruddigkeit_2012, H_se_2017}.
As an alternative to detailed quantum chemistry calculations, fully empirical machine learning (ML) predictions offer calculation times nearly six orders of magnitude faster than DFT ($O(10^{-3}s)$ for ML, $O(10^3s)$ for DFT on approximately 30 heavy atom molecules).
Machine learning approaches have recently been effective in reproducing DFT results given sufficient training data \citep{Faber:2017cs} and therefore offer an opportunity to efficiently screen much larger libraries of compounds without further reduction in chemical fidelity.

Developing ML pipelines for molecular property prediction often involves encoding variable-sized molecules as a finite-dimensional vector.
Traditional approaches use group contribution methods, molecular fingerprints, and molecular descriptors to convert molecular structures into a suitable input for dense neural networks or other ML models \citep{Katritzky:2010cp, Brown:2006gb, Maranas:1996hq, Pilania:2013di, St_John_2017, Das:2018gz, Janet_2017}.
However, hand-engineered molecular features may not sufficiently capture all the variability present in the space of chemically feasible compounds.
Neural network architectures that operate directly on graph-valued inputs have been developed \citep{Duvenaud2015}, allowing `end-to-end' learning on molecular space.
In this approach, models simultaneously learn both how to extract appropriate features as well as how to use these features to make accurate predictions.
End-to-end learning techniques have supplanted traditional methods in image recognition and computer translation, similar applications where determining a suitable fixed-size numerical representation of the input data is difficult.

A number of approaches for end-to-end learning on molecules have recently been unified into a single theoretical framework known as Message Passing Neural Networks (MPNNs), and even more recently as Graph Networks \citep{Gilmer2017, Battaglia_2018}.
In MPNNs, predictions are generated from input graphs with node and edge features.
The network comprises a sequence of layers, including a number of \emph{message passing} layers and a \emph{readout} layer.
In the message passing layers, node-level state vectors are updated according to graph's connectivity and the current states of neighboring nodes.
Following a number of message passing layers, the readout layer generates a single graph-level vector from node-level states.
These networks have demonstrated best-in-class performance on all properties in the QM9 computational dataset, a benchmark dataset for molecular property prediction consisting of DFT-optimized 3D coordinates and energies for $134,000$ molecules with nine or fewer heavy atoms \citep{Ramakrishnan_2014}.
Further modifications of the MPNN framework have demonstrated even higher accuracies \citep{NIPS2017_6700, Sch_tt_2018, Jorgensen_2018}.
However, both \citet{Gilmer2017} and more recent studies have noted that optimized, equilibrium 3D molecular geometries were required to achieve optimal accuracy on the QM9 dataset.
Since obtaining minimum-energy atomic coordinates is a numerically intensive task, this requirement is limiting for applications in high-throughput chemical screening - particularly for molecules with a large number of atoms.

While effective, deep learning requires large amounts of data in order to learn appropriate feature representations \citep{LeCun_2015}.
However, many applications of deep learning have benefited from \emph{transfer learning}, where weights from a neural network trained on a large dataset are used to initialize weights for a related task with limited data \citep{Yosinski2014}.
In this way, the model's ability to extract useful features from inputs - learned from the larger dataset - is transferred to the new regression task, improving predictive accuracy with fewer training samples.
In the molecular space, previous studies have shown that models are able to successfully predict molecules outside their training set \citep{Smith:2017jd, Yao_2018}, improve their predictive accuracy with additional training on molecules from a different distribution than the prediction target \citep{Sch_tt_2017}, and estimate non-equilibrium atomic energies at a higher level of theory by pretraining networks on lower-level calculations \citep{Smith_2018}.

In this study, we apply a MPNN to a newly developed computational dataset of $91,000$ molecules with optoelectronic calculations for organic photovoltaic (OPV) applications.
For OPV applications, single-molecule electronic properties play a role in determining overall device efficiency \citep{Oosterhout_2015, Scharber_2006, Li_2018}, and the search space of molecular structures is sufficiently large that experimental exploration is impractical.
Machine learning approaches have previously been used to predict the properties of candidate OPV materials \citep{Kanal_2013, Pyzer_Knapp_2015, J_rgensen_2018}, and a recent study demonstrated that a gap still exists between models that consider XYZ coordinates and those based only on SMILES strings \citep{J_rgensen_2018}.
While chemical structures of candidate molecules can be rapidly enumerated (referred to as a molecule's 2D geometry), calculating atomic positions at a high level of theory is computationally prohibitive when screening millions of possible molecules.
We therefore design a ML pipeline to predict optoelectronic properties (\emph{e.g.} $\epsilon_{\mathrm{HOMO}}$, $\epsilon_{\mathrm{LUMO}}$, optical excitation energy) directly from a molecule's 2D structure, without requiring 3D optimization using DFT.
We demonstrate that for the types of molecules considered in this study, MPNNs trained without explicit spatial information are capable of approaching chemical accuracy, and show nearly equivalent performance to models trained with spatial information.
Moreover, we show that weights from models trained on one DFT functional are able to improve performance on an alternative DFT functional with limited training data, even when the two target properties are poorly correlated.
This application demonstrates that high-throughput screening of molecular libraries (in the millions of molecules) can be accomplished at chemical accuracy quickly with machine learning methods without the computational burden of DFT structure optimization.
Additionally, these results indicate that the best neural network architectures trained on existing small-molecule quantum chemical datasets may not be optimal when molecular sizes increase.
We therefore make the newly developed OPV dataset considered in this work (with both 2D and 3D structures) publicly available for future graph network architecture development.

\section{Methods}

\subsection{Dataset preparation}

The database considered in this study contains calculations performed with several DFT functionals and basis sets (denoted functional/basis below) using the Gaussian 09 electronic structure package with default settings \citep{Gaussian2009}.
A web interface to the database is available at \texttt{[anonymized]}. 
The structures consist of combinations of building blocks, largely single and multi-ring heterocycles commonly found in OPV applications \citep{Scharber_2006, Oosterhout_2015, Lopez_2016}.
The database is primarily focused on quantifying the behavior of polymer systems, and therefore calculations were performed at a range of oligomer lengths to extrapolate to behavior at the polymer limit \citep{Larsen_2016}.
Two datasets were extracted from the database by selecting entries performed with the two functional/basis combinations with the greatest number of calculations, B3LYP/6-31g(d) and CAM-B3LYP/6-31g.
Each dataset consists of monomer structures, with or without 3D structural information, and associated DFT-calculated optoelectronic properties.
Molecular structures were encoded using SMILES strings \citep{Weininger_1988}, optimized 3D coordinates (when used) were stored in SDF files.
The specific electronic properties we predict are: the energy of highest occupied molecular orbital for the monomer ($\epsilon_{\mathrm{HOMO}}$); the lowest unoccupied molecular orbital of the monomer ($\epsilon_{\mathrm{LUMO}}$); the first excitation energy of the monomer calculated with time-dependent DFT (gap); the spectral overlap (integrated overlap between the optical absorption spectrum of a dimer and the AM1.5 solar spectrum); 
In addition to these properties, we also predict electronic properties that have been extrapolated to the polymer limit, including the polymer $\epsilon_{\mathrm{HOMO}}$, polymer $\epsilon_{\mathrm{LUMO}}$, polymer gap, and the optical $\epsilon_{\mathrm{LUMO}}$ (sum of the polymer $\epsilon_{\mathrm{HOMO}}$ and polymer gap).
In addition to polymers, the database also contains soluble small molecules for solution-processable OPV devices \citep{Lloyd_2007, van_der_Poll_2012}.
As these molecules are not polymerized, these entries lack information on extrapolated polymer electronics.
These entries were included in the training set, but excluded from the validation and test sets.


In order to screen a larger number of molecules, conformational sampling of each molecule was not performed; instead a single optimization was performed for each molecule or oligomer.
The primary B3LYP/6-31g(d) dataset consists of approximately $91,000$ molecules with unique SMILES strings, approximately $54,000$ of which contain polymer properties.
Of these $54,000$ with polymer properties, $5,000$ were randomly selected for each of the validation and test sets.
Transfer learning was examined with a secondary dataset consisting of results from the CAM-B3LYP/6-31g functional.
This dataset consists of approximately $32,000$ unique molecules, $17,000$ of which contain polymer results.
From the $17,000$ with polymer properties, $2,000$ were selected for the validation and test sets.
For both datasets, training data were randomly selected from the remainder small molecule and monomer results.
Prior to prediction, each property is scaled to have zero median and unit inner quartile range (followed by an inverse transformation after prediction).

Determining an appropriate optimal (or target) error rate that is representative of a best-case validation loss is an important step in optimizing the hyperparameters of a ML pipeline.
In previous studies, target errors were determined based on estimated experimental chemical accuracies for each of the regression tasks \citep{Faber:2017cs, Gilmer2017}.
However, since many of these parameters are not directly measurable experimentally, we sought to determine a target error directly from the data.
We therefore used calculation results from conformational isomers: molecules with identical connectivity but different 3D structure.
Due to the size of the considered molecules, energy minimization routines can often converge to different lowest-energy states, with slightly altered optoelectronic properties.
Since our model only considers atomic connectivity, it cannot distinguish between conformational isomers and predictions for molecules with identical SMILES strings will yield identical predictions.
By iterating over all pairs of conformers in the dataset, we calculate a mean absolute error (MAE) to establish a representative lower limit for predictive accuracy for a model that does not consider 3D atom positions.
These optimal errors are presented in Table~\ref{tbl:table1}.

\subsection{Message Passing Architecture}

The molecules considered in this study and used as building blocks for OPV polymers are relatively large, with a maximum size of 201 atoms and 424 bonds (including explicit hydrogens).
Inputs to the neural network are generated from the molecules' SMILES strings, and consists of discrete node types, edge types, and connectivity matrix.
Atoms are categorized into discrete types based on their atomic symbol, degree of bonding, and whether or not they are present in an aromatic ring.
Bonds are similarly categorized into discrete types based on their type (single, double, triple, or aromatic), conjugation, presence in a ring, and the atom symbols of the two participating atoms.

\begin{figure*}
\centering
\includegraphics[width=\textwidth]{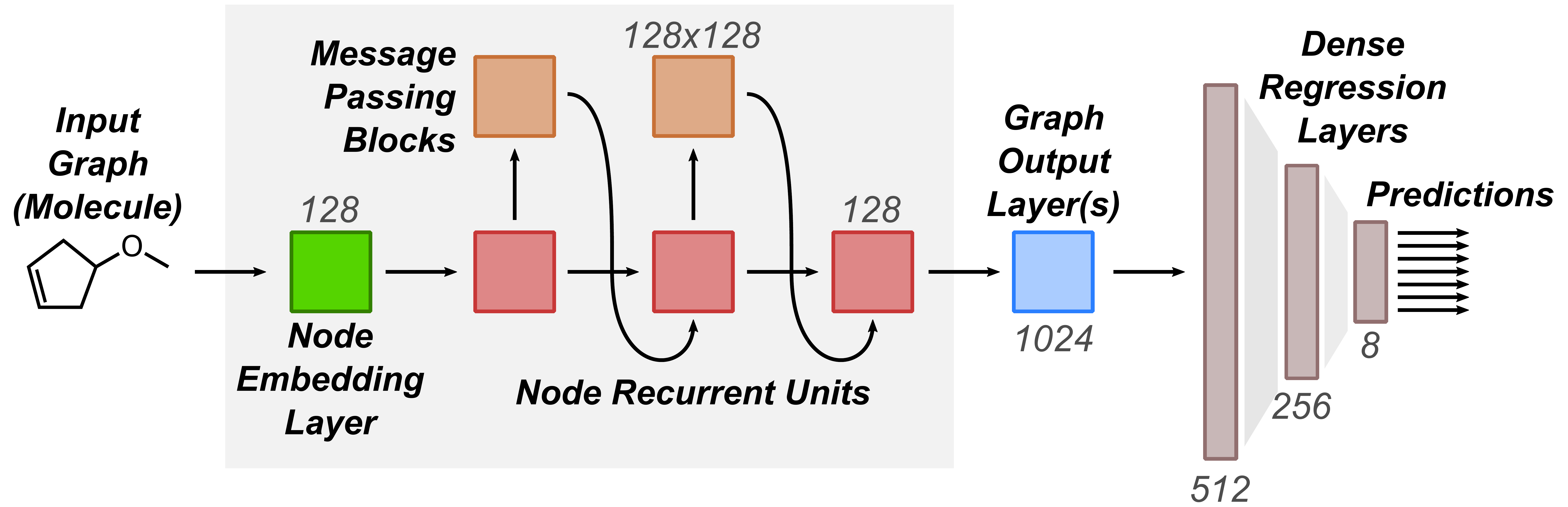}
\caption{\textbf{Schematic of the message passing framework for 2D structures.}
Input molecules are labelled according to their atom and bond types.
Atom embedding layers are used to initialize the weights of the message passing layers.
Molecule-level feature vectors are generated through an output layer which pools all atoms through summation, which is then passed to a series of dense layers to generate a final prediction.
Dimensions of each layer for the multi-task model are shown in gray.
For single-task models all dimensions are identical except for the final output layer, which has dimension 1.
}\label{fig:graph_learning}
\end{figure*}

A schematic of the neural network is shown in Figure~\ref{fig:graph_learning}.
The \emph{message passing} step was implemented using the matrix multiplication method \citep{Li2016, Gilmer2017}, where messages $m$ are passed between neighboring atoms, $$m_v^{t+1} = \sum_{w \in N(v)} A_{e_{vw}} h_w^t,$$ where $v$ is the node index, $N(v)$ are the neighboring nodes, $e_{vw}$ is the bond type, $h_v^t$ is the feature vector for node $v$ at step $t$, and $A_{e_{vw}}$ is a learned weight matrix for each bond type.

The \emph{update} step was implemented as a gated recurrent unit block \citep{Gilmer2017}, $$h_v^{t+1} = \mathrm{GRU}(h_v^t, m_v^{t+1}).$$ Initial atom embeddings, $h_v^0$, are initialized randomly for each atom class and learned as additional model parameters.
The dimension of the atom state was chosen to be $128$, with $M=3$ message-passing layers.
The readout function used was similar to the one used by \citet{Duvenaud2015}, but uses only the final hidden state of the recurrent atom unit to generate a whole-graph feature vector $\hat{y}$: $$\hat{y} = \sum_{v \in G} \sigma(W h_v^M),$$ where $W$ is a learned weight matrix.
The dimension of $\hat{y}$ was chosen to be $1024$.
This summed fingerprint is then passed through a series of two fully connected layers with batch normalization and ReLU activation functions (dimensions $512$ and $256$, respectively), before being passed to an output layer corresponding to each property prediction.

When 3D molecular geometries were considered, the SchNet structure with edge updates from \citet{Jorgensen_2018} was used.
A nearest-neighbor cutoff of $48$ was used to determine the connectivity matrix of passed messages.
The dimension of the atom hidden state was chosen as $C = 64$, and separate models were trained for each of the eight target properties.
As the targets are mainly orbital energies, we similarly use a average in the readout function.
SchNet-like models were trained with the ADAM optimizer with an initial learning rate of $1E-4$, and a decay rate of $1E-5$ per epoch.
The models used a batch size of 32 and were trained for 500 epochs.

\subsection{Software}

Message passing operations were implemented using Keras and Tensorflow.
Scikit-learn was used to scale the prediction targets, and rdkit was used to encode the atoms and bonds as integer classes.
A python library used to implement the MPNNs described in this study is available on Github (\url{github.com/nrel/nfp}) and installable via \texttt{pip}.
All datasets, model scripts, and trained model weights for the models described in Table~\ref{tbl:table1} are available at \url{https://cscdata.nrel.gov/#/datasets/ad5d2c9a-af0a-4d72-b943-1e433d5750d6}.

\subsection{Hyperparameter optimization}

For the 2D model, model sizes (atom vector dimension, molecule vector dimension, number and size of dense layers) were increased until training errors fell below the target optimal error rate while the model still fit on single GPU (Tesla K80) with a batch size of 100.
Models were optimized using the ADAM optimizer.
Learning rates were varied between 1E-2 and 1E-5, with 1E-3 yielding the best result.
Explicit learning rate decay was also noticed to improve optimization, a decay value of 2E-6 each epoch was used.
Models were trained for 500 epochs.
Methods for explicit regularization, including dropout and $l_2$ schemes were tried, but did not decrease the validation loss.
All models (including refitting weights during transfer learning) used early stopping by evaluating the validation loss every 10 epochs and using the model which yielded the lowest validation loss.

\section{Results}

\subsection{Prediction performance on B3LYP/6-31g(d) results}

The largest database consists of calculations performed at the B3LYP/6-31g(d) level of theory.
By comparing calculation results for molecules with identical SMILES strings but different 3D geometries, a baseline error rate was established for models that only considered SMILES strings (2D features) as inputs.
This error rate was relatively low: for $\epsilon_{\mathrm{HOMO}}$, the mean absolute error (MAE) between pairs of conformers was $28.0$ meV, lower than both the target "chemical accuracy" of $43$ meV used in \citet{Faber:2017cs} and the MAE reached by the current best-performing model on the QM9 dataset, $36.7$ meV \cite{Jorgensen_2018}.

Two strategies were used to train models using only 2D coordinates.
First, a series of models were trained for each property (Table~\ref{tbl:table1}, ``2D, single-task'').
These models were capable of closely matching DFT results, with MAEs in orbital energies approximately $10$ meV higher than the calculated optimal error.
These errors, $32.1$ meV for $\epsilon_{\mathrm{HOMO}}$, are lower than state-of-the-art models on the QM9 dataset, suggesting 2D connectivity is sufficient to specify molecular properties for these types of molecules.
Next, a single model was trained to simultaneously predict all eight target properties (``2D, multi-task'').
This model greatly improves prediction speed while demonstrating similar error rates to the single-task models.

\begin{figure*}[t!]
\centering
\includegraphics[width=\textwidth]{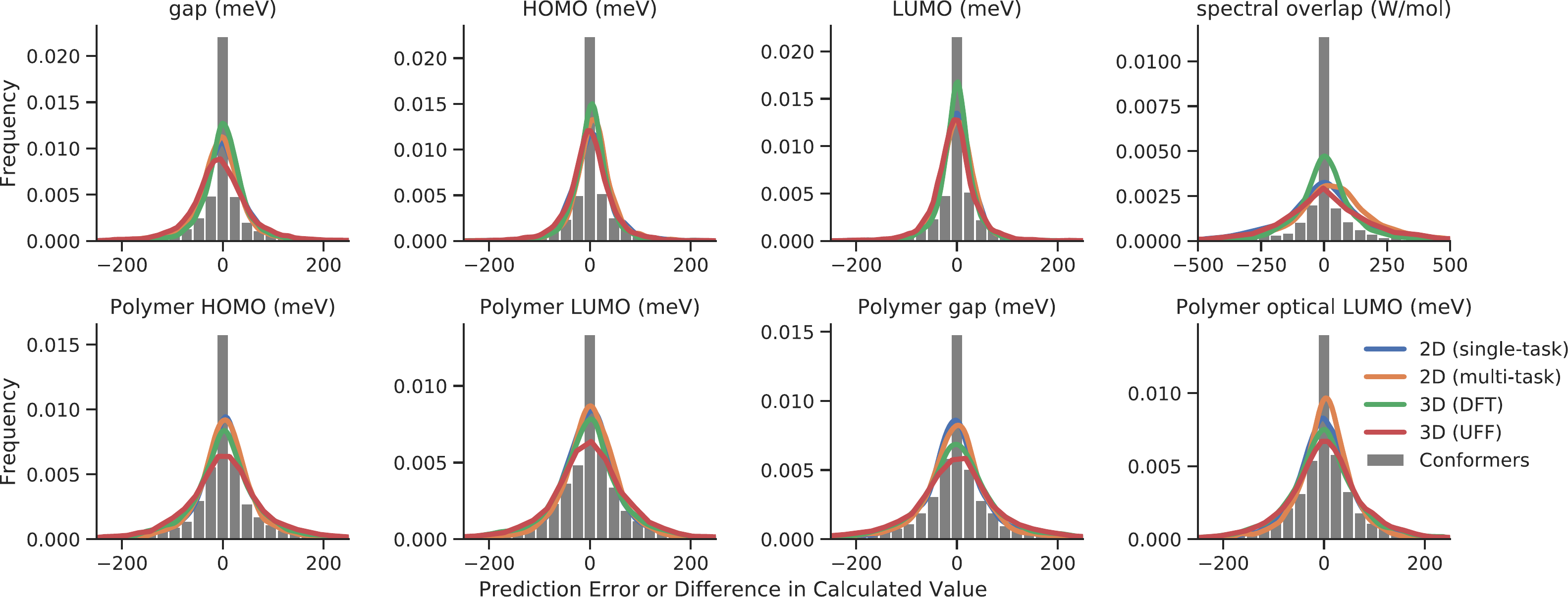}
\caption{\textbf{Distributions in prediction error for held-out data.}
	Distributions in prediction errors for test-set molecules from each model summarized in Table~\ref{tbl:table1}.
	Histograms in differences in calculated values between pairs of conformational isomers are shown in gray.
	Lines represent kernel density estimates for prediction errors from each model.
}\label{fig:error_hists}
\end{figure*}

For comparison, models were also trained using DFT-optimized 3D coordinates.
The MPNN structure of these models were adapted from that of \citet{Jorgensen_2018}, and a single model was trained for each target property.
Resulting error distributions were similar to those of models trained on only 2D coordinates (Table~\ref{tbl:table1}, ``3D, DFT''; Figure~\ref{fig:error_hists}).
The similarity in error distributions between models which consider 3D and 2D further indicates that for the molecules considered in this database, 2D structural information is sufficient to specify optoelectronic properties.
Errors for the 3D model were smaller for monomer and dimer properties (gap, $\epsilon_{\mathrm{HOMO}}$, $\epsilon_{\mathrm{LUMO}}$, spectral overlap), while slightly larger for extrapolated polymer properties.
This effect may suggest that polymer properties are less dependent on the monomer's precise 3D configuration.

Approximate 3D coordinates can be computed rapidly using empirical force fields, for instance the UFF force field \citep{Rappe_1992}.
Molecules in the dataset were re-optimized using the UFF force field, in order to determine approximate 3D coordinates at a much lower computational cost.
Models were then re-trained using these approximated geometries.
The resulting prediction accuracies were worse than even the 2D models, indicating that using poor-quality molecular geometries gives worse results than omitting 3D features (Table~\ref{tbl:table1}, ``3D, UFF'').

\begin{table}[tb]
\centering

\caption{\label{tbl:table1}
\textbf{Mean absolute errors (MAEs) for test set predictions for models trained on B3LYP/6-31g(d) results.}
The conformers column reports MAE between calculations for pairs of conformational isomers, representing an optimal error rate for models trained on 2D coordinates.
Distributions of prediction errors are shown in Figure~\ref{fig:error_hists}.}

\begin{tabular}{llllll}
\toprule
 & & \multicolumn{2}{c}{2D} & \multicolumn{2}{c}{3D} \\
\cmidrule(lr){3-4}
\cmidrule(lr){5-6}

\textbf{B3LYP/6-31g(d)}                 & \emph{Conformers} & single-task & multi-task & DFT  & UFF\\\midrule
gap                                     & \emph{28.0 meV}   & 36.9        & 35.4       & 32.7 & 45.1 \\
$\epsilon_{\mathrm{HOMO}}$              & \emph{22.0 meV}   & 32.1        & 29.4       & 27.0 & 33.1 \\
$\epsilon_{\mathrm{LUMO}}$              & \emph{25.5 meV}   & 27.9        & 29.2       & 24.8 & 33.9 \\
spectral overlap                        & \emph{81.3 W/mol} & 149.3       & 149.2      & 96.6 & 170.0 \\
Polymer $\epsilon_{\mathrm{HOMO}}$      & \emph{37.4 meV}   & 49.1        & 47.4       & 56.9 & 64.8 \\
Polymer $\epsilon_{\mathrm{LUMO}}$      & \emph{45.0 meV}   & 47.8        & 46.8       & 56.8 & 63.0 \\
Polymer gap                             & \emph{46.3 meV}   & 57.1        & 56.3       & 69.8 & 74.3 \\
Pol. optical $\epsilon_{\mathrm{LUMO}}$ & \emph{42.6 meV}   & 47.8        & 43.9       & 57.2 & 60.2 \\

\bottomrule
\end{tabular}
\end{table}

We next explored the effect of training set size on prediction accuracy for models trained on 2D structures.
Repeated optimizations of the multi-task model were performed with sub-sampled training data with the validation set, test set, and model architecture held constant across all experiments.
As expected, additional training data causes out-of-sample predictive performance to improve, shown in Figure~\ref{fig:learning_curves}A.
The model's accuracy asymptotically approaches the optimal error rate at the largest training set sizes.

\subsection{Transfer learning to an alternate DFT functional}

Finally, we examined whether the molecular representations learned from the large-scale B3LYP/6-31g(d) dataset improved predictive performance on a related regression.
End-to-end learning models perform two tasks: they extract salient features from the input data and recombine these features to generate a prediction.
Inside the network, higher level representations of the data are produced by subsequent layers before ultimately leading to a predicted value.  
Transferring weights to a new model from a model trained on a closely correlated target can therefore preserve much of the logic and higher-level representations of the previous model. 
However, even transferring weights from a poorly correlated target can aid models by preserving low-level features useful for both targets.

\begin{figure}[t!]
\centering
\includegraphics[width=\columnwidth]{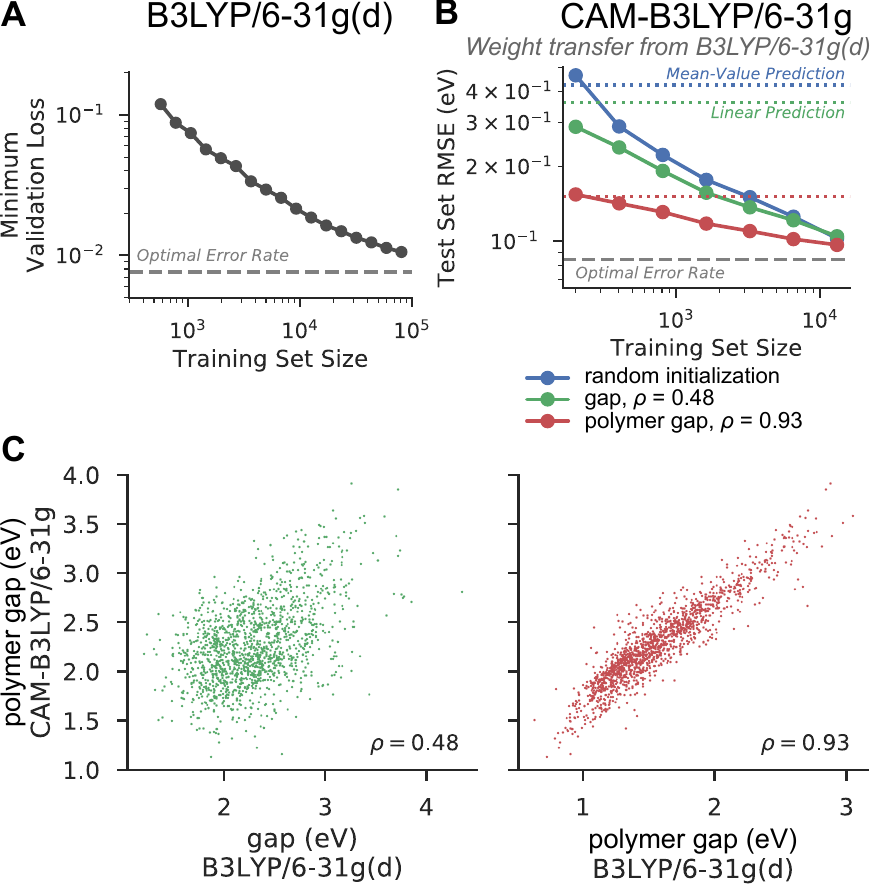}
\caption{\textbf{Effect of training set size on predictive performance} (A) Training on B3LYP/6-31g(d).
Models gradually approach the optimal error rate as training set size increases.
(B) Transfer learning to predict polymer band gap calculated with CAM-B3LYP/6-31g.
For each model, performance is compared to both the optimal error rate and an estimated upper error bound based on a simple linear model (dotted lines).
(C) Illustration of the similarity between old and new prediction tasks considered during transfer learning. Plot of CAM-B3LYP/6-31g polymer band gap (new) versus the single-target properties used for pretraining: monomer band gap (left) and polymer band gap (right).
Points represent molecules with results calculated via both functionals.}\label{fig:learning_curves}
\end{figure}

To test the effectiveness of transfer learning with the proposed MPNN structure, a second, smaller dataset of polymer band gap values calculated using the CAM-B3LYP/6-31g functional was used as a benchmark task.
Two models trained on B3LYP/6-31g(d) data were used to initialize the weights for a new polymer band gap prediction model: first, a model trained on the same parameter calculated via B3LYP/6-31g(d), and, as a more difficult example, a model trained on the B3LYP/6-31g(d) monomer band gap.
Correlation coefficients were used as a measure of the similarity between the old and new prediction targets.
The correlation coefficients between the CAM-B3LYP/6-31g polymer band gap and B3LYP polymer and monomer band gaps were 0.93 and 0.48, respectively, for molecules present in both the CAM-B3LYP/6-31g and B3LYP/6-31g(d) datasets Figure~\ref{fig:learning_curves}C.

Test and validation sets of 2,000 polymer species were reserved, and the remaining data was sub-divided into training sets of increasing size.
All transfer learning strategies were compared against a reference model with random weight initialization for all layers (i.e., no transfer learning).
The results of all model predictions on the test set are shown in Figure~\ref{fig:learning_curves}B.
For each model, performance is compared to an estimated upper-bound error.
For the reference model this error was equal to the data's standard deviation, assuming a worst-case model would always predict the mean value of the prediction target.
For the models with transferred weights, upper-bound errors were found assuming new targets were calculated by linearly transforming the old prediction target to best match the new target.
The root mean squared error (RMSE) for these two base-case models were calculated as $360$ meV and $150$ meV for B3LYP/6-31g(d) monomer band gap and polymer band gap, respectively.
For models with weight transfer, performance superior to these estimated upper error limits indicates that the model has retained the ability to extract and process salient features of the molecules related to the new prediction target - rather than simply recalling and rescaling the previously learned output.

For very small training set sizes (on the order of 200 molecules), models performed near the estimated upper error bound with the notable exception of the model with weakly correlated transferred weights, which had a substantially lower test set error than expected.
This result demonstrates that pretraining models on even slightly related prediction targets could likely improve out-of-sample prediction accuracy when the available data is limited by allowing the MPNNs to learn useful molecular features.
As the available training data is increased, both models with transferred weights demonstrate a concomitant decrease in their test set error below their estimated upper bound error.
In particular, the model with weights transferred from the strongly correlated task shows superior performance at all training set sizes, requiring nearly an order of magnitude less data to reach RMSE values of $100$ meV.
At the largest training set sizes all three models approach the optimal error rate (estimated through conformers with duplicated SMILES strings), indicating that knowledge encapsulated in transferred weights is eventually replaced with knowledge gained through the new training data.

\section{Conclusions}

In this study, we have demonstrated near-equivalent prediction accuracies from both 2D and 3D structural features in MPNN architectures, both of which closely approach the estimated 2D lower-bound error from conformational optimization.
While studies on the QM9 dataset have shown that 3D coordinates are required for accurate predictions, using these data as inputs mandates that full DFT calculations still be performed for each molecule.
The necessity of 3D coordinates for the QM9 dataset might be explained by the substantially smaller molecules considered ($\le 29$ atoms, including hydrogens) when compared with our newly generated OPV database ($\le 201$ atoms).
Additionally, since they are exhaustively generated according to computational rules, molecules in QM9 frequently contain complex structural features that might only be captured through the explicit use of 3D coordinates.
Our new public database might therefore serve as a more representative molecular learning benchmark for electronic structure calculations.

We have shown that a deep neural network pretrained on one DFT functional was able to improve predictive performance on a related DFT functional, especially in the case of limited data.
This performance improvement is dependent on the correlation between tasks, but even weights transferred from a network trained on a weakly correlated task were able to improve accuracy.
These results help to confirm the immense value of machine learning approaches in scientific domains both to increase the fidelity of DFT simulations and to augment them, allowing for high throughput screening and guided search.
Future work will therefore explore the ability of pretrained neural networks to improve prediction accuracy on experimental data and other important targets with limited available data.

\section*{Acknowledgements}

This work was supported by the U.S.  Department of Energy under Contract No.  DE-AC36-08GO28308 with Alliance for Sustainable Energy, LLC, the Manager and Operator of the National Renewable Energy Laboratory.
Funding was provided by U.S.  Department of Energy, Office of Bioenergy Technologies (DOE-BETO) under the Performance Advantaged Bioproducts working group.
The views and opinions of the authors expressed herein do not necessarily state or reflect those of the United States Government or any agency thereof.
Neither the United States Government nor any agency thereof, nor any of their employees, makes any warranty, expressed or implied, or assumes any legal liability or responsibility for the accuracy, completeness, or usefulness of any information, apparatus, product, or process disclosed, or represents that its use would not infringe privately owned rights.

\nocite{*}
\bibliography{library.bib}

\begin{thebibliography}{40}%
\makeatletter
\providecommand \@ifxundefined [1]{%
 \@ifx{#1\undefined}
}%
\providecommand \@ifnum [1]{%
 \ifnum #1\expandafter \@firstoftwo
 \else \expandafter \@secondoftwo
 \fi
}%
\providecommand \@ifx [1]{%
 \ifx #1\expandafter \@firstoftwo
 \else \expandafter \@secondoftwo
 \fi
}%
\providecommand \natexlab [1]{#1}%
\providecommand \enquote  [1]{``#1''}%
\providecommand \bibnamefont  [1]{#1}%
\providecommand \bibfnamefont [1]{#1}%
\providecommand \citenamefont [1]{#1}%
\providecommand \href@noop [0]{\@secondoftwo}%
\providecommand \href [0]{\begingroup \@sanitize@url \@href}%
\providecommand \@href[1]{\@@startlink{#1}\@@href}%
\providecommand \@@href[1]{\endgroup#1\@@endlink}%
\providecommand \@sanitize@url [0]{\catcode `\\12\catcode `\$12\catcode
  `\&12\catcode `\#12\catcode `\^12\catcode `\_12\catcode `\%12\relax}%
\providecommand \@@startlink[1]{}%
\providecommand \@@endlink[0]{}%
\providecommand \url  [0]{\begingroup\@sanitize@url \@url }%
\providecommand \@url [1]{\endgroup\@href {#1}{\urlprefix }}%
\providecommand \urlprefix  [0]{URL }%
\providecommand \Eprint [0]{\href }%
\providecommand \doibase [0]{http://dx.doi.org/}%
\providecommand \selectlanguage [0]{\@gobble}%
\providecommand \bibinfo  [0]{\@secondoftwo}%
\providecommand \bibfield  [0]{\@secondoftwo}%
\providecommand \translation [1]{[#1]}%
\providecommand \BibitemOpen [0]{}%
\providecommand \bibitemStop [0]{}%
\providecommand \bibitemNoStop [0]{.\EOS\space}%
\providecommand \EOS [0]{\spacefactor3000\relax}%
\providecommand \BibitemShut  [1]{\csname bibitem#1\endcsname}%
\let\auto@bib@innerbib\@empty
\bibitem [{\citenamefont {Jain}\ \emph {et~al.}(2013)\citenamefont {Jain},
  \citenamefont {Ong}, \citenamefont {Hautier}, \citenamefont {Chen},
  \citenamefont {Richards}, \citenamefont {Dacek}, \citenamefont {Cholia},
  \citenamefont {Gunter}, \citenamefont {Skinner}, \citenamefont {Ceder},\ and\
  \citenamefont {Persson}}]{Jain_2013}%
  \BibitemOpen
  \bibfield  {author} {\bibinfo {author} {\bibfnamefont {A.}~\bibnamefont
  {Jain}}, \bibinfo {author} {\bibfnamefont {S.~P.}\ \bibnamefont {Ong}},
  \bibinfo {author} {\bibfnamefont {G.}~\bibnamefont {Hautier}}, \bibinfo
  {author} {\bibfnamefont {W.}~\bibnamefont {Chen}}, \bibinfo {author}
  {\bibfnamefont {W.~D.}\ \bibnamefont {Richards}}, \bibinfo {author}
  {\bibfnamefont {S.}~\bibnamefont {Dacek}}, \bibinfo {author} {\bibfnamefont
  {S.}~\bibnamefont {Cholia}}, \bibinfo {author} {\bibfnamefont
  {D.}~\bibnamefont {Gunter}}, \bibinfo {author} {\bibfnamefont
  {D.}~\bibnamefont {Skinner}}, \bibinfo {author} {\bibfnamefont
  {G.}~\bibnamefont {Ceder}}, \ and\ \bibinfo {author} {\bibfnamefont {K.~A.}\
  \bibnamefont {Persson}},\ }\bibfield  {title} {\enquote {\bibinfo {title}
  {Commentary: The materials project: A materials genome approach to
  accelerating materials innovation},}\ }\href {\doibase 10.1063/1.4812323}
  {\bibfield  {journal} {\bibinfo  {journal} {APL Mater}\ }\textbf {\bibinfo
  {volume} {1}},\ \bibinfo {pages} {011002} (\bibinfo {year}
  {2013})}\BibitemShut {NoStop}%
\bibitem [{\citenamefont {Lopez}\ \emph {et~al.}(2016)\citenamefont {Lopez},
  \citenamefont {Pyzer-Knapp}, \citenamefont {Simm}, \citenamefont {Lutzow},
  \citenamefont {Li}, \citenamefont {Seress}, \citenamefont {Hachmann},\ and\
  \citenamefont {Aspuru-Guzik}}]{Lopez_2016}%
  \BibitemOpen
  \bibfield  {author} {\bibinfo {author} {\bibfnamefont {S.~A.}\ \bibnamefont
  {Lopez}}, \bibinfo {author} {\bibfnamefont {E.~O.}\ \bibnamefont
  {Pyzer-Knapp}}, \bibinfo {author} {\bibfnamefont {G.~N.}\ \bibnamefont
  {Simm}}, \bibinfo {author} {\bibfnamefont {T.}~\bibnamefont {Lutzow}},
  \bibinfo {author} {\bibfnamefont {K.}~\bibnamefont {Li}}, \bibinfo {author}
  {\bibfnamefont {L.~R.}\ \bibnamefont {Seress}}, \bibinfo {author}
  {\bibfnamefont {J.}~\bibnamefont {Hachmann}}, \ and\ \bibinfo {author}
  {\bibfnamefont {A.}~\bibnamefont {Aspuru-Guzik}},\ }\bibfield  {title}
  {\enquote {\bibinfo {title} {The harvard organic photovoltaic dataset},}\
  }\href {\doibase 10.1038/sdata.2016.86} {\bibfield  {journal} {\bibinfo
  {journal} {Sci Data}\ }\textbf {\bibinfo {volume} {3}},\ \bibinfo {pages}
  {160086} (\bibinfo {year} {2016})}\BibitemShut {NoStop}%
\bibitem [{\citenamefont {Huang}\ \emph {et~al.}(2017)\citenamefont {Huang},
  \citenamefont {Shang}, \citenamefont {Zhang},\ and\ \citenamefont
  {Liu}}]{Huang_2017}%
  \BibitemOpen
  \bibfield  {author} {\bibinfo {author} {\bibfnamefont {S.-D.}\ \bibnamefont
  {Huang}}, \bibinfo {author} {\bibfnamefont {C.}~\bibnamefont {Shang}},
  \bibinfo {author} {\bibfnamefont {X.-J.}\ \bibnamefont {Zhang}}, \ and\
  \bibinfo {author} {\bibfnamefont {Z.-P.}\ \bibnamefont {Liu}},\ }\bibfield
  {title} {\enquote {\bibinfo {title} {Material discovery by combining
  stochastic surface walking global optimization with a neural network},}\
  }\href {\doibase 10.1039/c7sc01459g} {\bibfield  {journal} {\bibinfo
  {journal} {Chem Sci}\ }\textbf {\bibinfo {volume} {8}},\ \bibinfo {pages}
  {6327–6337} (\bibinfo {year} {2017})}\BibitemShut {NoStop}%
\bibitem [{\citenamefont {Ruddigkeit}\ \emph {et~al.}(2012)\citenamefont
  {Ruddigkeit}, \citenamefont {van Deursen}, \citenamefont {Blum},\ and\
  \citenamefont {Reymond}}]{Ruddigkeit_2012}%
  \BibitemOpen
  \bibfield  {author} {\bibinfo {author} {\bibfnamefont {L.}~\bibnamefont
  {Ruddigkeit}}, \bibinfo {author} {\bibfnamefont {R.}~\bibnamefont {van
  Deursen}}, \bibinfo {author} {\bibfnamefont {L.~C.}\ \bibnamefont {Blum}}, \
  and\ \bibinfo {author} {\bibfnamefont {J.-L.}\ \bibnamefont {Reymond}},\
  }\bibfield  {title} {\enquote {\bibinfo {title} {Enumeration of 166 billion
  organic small molecules in the chemical universe database gdb-17},}\ }\href
  {\doibase 10.1021/ci300415d} {\bibfield  {journal} {\bibinfo  {journal} {J
  Chem Inf Model}\ }\textbf {\bibinfo {volume} {52}},\ \bibinfo {pages}
  {2864--2875} (\bibinfo {year} {2012})}\BibitemShut {NoStop}%
\bibitem [{\citenamefont {Häse}, \citenamefont {Kreisbeck},\ and\
  \citenamefont {Aspuru-Guzik}(2017)}]{H_se_2017}%
  \BibitemOpen
  \bibfield  {author} {\bibinfo {author} {\bibfnamefont {F.}~\bibnamefont
  {Häse}}, \bibinfo {author} {\bibfnamefont {C.}~\bibnamefont {Kreisbeck}}, \
  and\ \bibinfo {author} {\bibfnamefont {A.}~\bibnamefont {Aspuru-Guzik}},\
  }\bibfield  {title} {\enquote {\bibinfo {title} {Machine learning for quantum
  dynamics: deep learning of excitation energy transfer properties},}\ }\href
  {\doibase 10.1039/c7sc03542j} {\bibfield  {journal} {\bibinfo  {journal}
  {Chem Sci}\ }\textbf {\bibinfo {volume} {8}},\ \bibinfo {pages} {8419–8426}
  (\bibinfo {year} {2017})}\BibitemShut {NoStop}%
\bibitem [{\citenamefont {Faber}\ \emph {et~al.}(2017)\citenamefont {Faber},
  \citenamefont {Hutchison}, \citenamefont {Huang}, \citenamefont {Gilmer},
  \citenamefont {Schoenholz}, \citenamefont {Dahl}, \citenamefont {Vinyals},
  \citenamefont {Kearnes}, \citenamefont {Riley},\ and\ \citenamefont {{von
  Lilienfeld}}}]{Faber:2017cs}%
  \BibitemOpen
  \bibfield  {author} {\bibinfo {author} {\bibfnamefont {F.~A.}\ \bibnamefont
  {Faber}}, \bibinfo {author} {\bibfnamefont {L.}~\bibnamefont {Hutchison}},
  \bibinfo {author} {\bibfnamefont {B.}~\bibnamefont {Huang}}, \bibinfo
  {author} {\bibfnamefont {J.}~\bibnamefont {Gilmer}}, \bibinfo {author}
  {\bibfnamefont {S.~S.}\ \bibnamefont {Schoenholz}}, \bibinfo {author}
  {\bibfnamefont {G.~E.}\ \bibnamefont {Dahl}}, \bibinfo {author}
  {\bibfnamefont {O.}~\bibnamefont {Vinyals}}, \bibinfo {author} {\bibfnamefont
  {S.}~\bibnamefont {Kearnes}}, \bibinfo {author} {\bibfnamefont {P.~F.}\
  \bibnamefont {Riley}}, \ and\ \bibinfo {author} {\bibfnamefont {O.~A.}\
  \bibnamefont {{von Lilienfeld}}},\ }\bibfield  {title} {\enquote {\bibinfo
  {title} {Prediction errors of molecular machine learning models lower than
  hybrid {DFT} error},}\ }\href {\doibase 10.1021/acs.jctc.7b00577} {\bibfield
  {journal} {\bibinfo  {journal} {J Chem Theory Comput}\ }\textbf {\bibinfo
  {volume} {13}},\ \bibinfo {pages} {5255--5264} (\bibinfo {year}
  {2017})}\BibitemShut {NoStop}%
\bibitem [{\citenamefont {Katritzky}\ \emph {et~al.}(2010)\citenamefont
  {Katritzky}, \citenamefont {Kuanar}, \citenamefont {Slavov}, \citenamefont
  {Hall}, \citenamefont {Karelson}, \citenamefont {Kahn},\ and\ \citenamefont
  {Dobchev}}]{Katritzky:2010cp}%
  \BibitemOpen
  \bibfield  {author} {\bibinfo {author} {\bibfnamefont {A.~R.}\ \bibnamefont
  {Katritzky}}, \bibinfo {author} {\bibfnamefont {M.}~\bibnamefont {Kuanar}},
  \bibinfo {author} {\bibfnamefont {S.}~\bibnamefont {Slavov}}, \bibinfo
  {author} {\bibfnamefont {C.~D.}\ \bibnamefont {Hall}}, \bibinfo {author}
  {\bibfnamefont {M.}~\bibnamefont {Karelson}}, \bibinfo {author}
  {\bibfnamefont {I.}~\bibnamefont {Kahn}}, \ and\ \bibinfo {author}
  {\bibfnamefont {D.~A.}\ \bibnamefont {Dobchev}},\ }\bibfield  {title}
  {\enquote {\bibinfo {title} {Quantitative correlation of physical and
  chemical properties with chemical structure: Utility for prediction},}\
  }\href {\doibase 10.1021/cr900238d} {\bibfield  {journal} {\bibinfo
  {journal} {Chem Rev}\ }\textbf {\bibinfo {volume} {110}},\ \bibinfo {pages}
  {5714--5789} (\bibinfo {year} {2010})}\BibitemShut {NoStop}%
\bibitem [{\citenamefont {Brown}\ \emph {et~al.}(2006)\citenamefont {Brown},
  \citenamefont {Martin}, \citenamefont {Rintoul},\ and\ \citenamefont
  {Faulon}}]{Brown:2006gb}%
  \BibitemOpen
  \bibfield  {author} {\bibinfo {author} {\bibfnamefont {W.~M.}\ \bibnamefont
  {Brown}}, \bibinfo {author} {\bibfnamefont {S.}~\bibnamefont {Martin}},
  \bibinfo {author} {\bibfnamefont {M.~D.}\ \bibnamefont {Rintoul}}, \ and\
  \bibinfo {author} {\bibfnamefont {J.-L.}\ \bibnamefont {Faulon}},\ }\bibfield
   {title} {\enquote {\bibinfo {title} {Designing novel polymers with targeted
  properties using the signature molecular descriptor},}\ }\href {\doibase
  10.1021/ci0504521} {\bibfield  {journal} {\bibinfo  {journal} {J Chem Inf
  Model}\ }\textbf {\bibinfo {volume} {46}},\ \bibinfo {pages} {826--835}
  (\bibinfo {year} {2006})}\BibitemShut {NoStop}%
\bibitem [{\citenamefont {Maranas}(1996)}]{Maranas:1996hq}%
  \BibitemOpen
  \bibfield  {author} {\bibinfo {author} {\bibfnamefont {C.~D.}\ \bibnamefont
  {Maranas}},\ }\bibfield  {title} {\enquote {\bibinfo {title} {Optimal
  computer-aided molecular design:~ a polymer design case study},}\ }\href
  {\doibase 10.1021/ie960096z} {\bibfield  {journal} {\bibinfo  {journal} {Ind
  Eng Chem Res}\ }\textbf {\bibinfo {volume} {35}},\ \bibinfo {pages}
  {3403--3414} (\bibinfo {year} {1996})}\BibitemShut {NoStop}%
\bibitem [{\citenamefont {Pilania}\ \emph {et~al.}(2013)\citenamefont
  {Pilania}, \citenamefont {Wang}, \citenamefont {Jiang}, \citenamefont
  {Rajasekaran},\ and\ \citenamefont {Ramprasad}}]{Pilania:2013di}%
  \BibitemOpen
  \bibfield  {author} {\bibinfo {author} {\bibfnamefont {G.}~\bibnamefont
  {Pilania}}, \bibinfo {author} {\bibfnamefont {C.}~\bibnamefont {Wang}},
  \bibinfo {author} {\bibfnamefont {X.}~\bibnamefont {Jiang}}, \bibinfo
  {author} {\bibfnamefont {S.}~\bibnamefont {Rajasekaran}}, \ and\ \bibinfo
  {author} {\bibfnamefont {R.}~\bibnamefont {Ramprasad}},\ }\bibfield  {title}
  {\enquote {\bibinfo {title} {Accelerating materials property predictions
  using machine learning},}\ }\href {\doibase 10.1038/srep02810} {\bibfield
  {journal} {\bibinfo  {journal} {Sci Rep}\ }\textbf {\bibinfo {volume} {3}}
  (\bibinfo {year} {2013}),\ 10.1038/srep02810}\BibitemShut {NoStop}%
\bibitem [{\citenamefont {St.~John}\ \emph {et~al.}(2017)\citenamefont
  {St.~John}, \citenamefont {Kairys}, \citenamefont {Das}, \citenamefont
  {McEnally}, \citenamefont {Pfefferle}, \citenamefont {Robichaud},
  \citenamefont {Nimlos}, \citenamefont {Zigler}, \citenamefont {McCormick},
  \citenamefont {Foust}, \citenamefont {Bomble},\ and\ \citenamefont
  {Kim}}]{St_John_2017}%
  \BibitemOpen
  \bibfield  {author} {\bibinfo {author} {\bibfnamefont {P.~C.}\ \bibnamefont
  {St.~John}}, \bibinfo {author} {\bibfnamefont {P.}~\bibnamefont {Kairys}},
  \bibinfo {author} {\bibfnamefont {D.~D.}\ \bibnamefont {Das}}, \bibinfo
  {author} {\bibfnamefont {C.~S.}\ \bibnamefont {McEnally}}, \bibinfo {author}
  {\bibfnamefont {L.~D.}\ \bibnamefont {Pfefferle}}, \bibinfo {author}
  {\bibfnamefont {D.~J.}\ \bibnamefont {Robichaud}}, \bibinfo {author}
  {\bibfnamefont {M.~R.}\ \bibnamefont {Nimlos}}, \bibinfo {author}
  {\bibfnamefont {B.~T.}\ \bibnamefont {Zigler}}, \bibinfo {author}
  {\bibfnamefont {R.~L.}\ \bibnamefont {McCormick}}, \bibinfo {author}
  {\bibfnamefont {T.~D.}\ \bibnamefont {Foust}}, \bibinfo {author}
  {\bibfnamefont {Y.~J.}\ \bibnamefont {Bomble}}, \ and\ \bibinfo {author}
  {\bibfnamefont {S.}~\bibnamefont {Kim}},\ }\bibfield  {title} {\enquote
  {\bibinfo {title} {A quantitative model for the prediction of sooting
  tendency from molecular structure},}\ }\href {\doibase
  10.1021/acs.energyfuels.7b00616} {\bibfield  {journal} {\bibinfo  {journal}
  {Energy Fuels}\ }\textbf {\bibinfo {volume} {31}},\ \bibinfo {pages}
  {9983--9990} (\bibinfo {year} {2017})}\BibitemShut {NoStop}%
\bibitem [{\citenamefont {Das}\ \emph {et~al.}(2018)\citenamefont {Das},
  \citenamefont {John}, \citenamefont {McEnally}, \citenamefont {Kim},\ and\
  \citenamefont {Pfefferle}}]{Das:2018gz}%
  \BibitemOpen
  \bibfield  {author} {\bibinfo {author} {\bibfnamefont {D.~D.}\ \bibnamefont
  {Das}}, \bibinfo {author} {\bibfnamefont {P.~C.~S.}\ \bibnamefont {John}},
  \bibinfo {author} {\bibfnamefont {C.~S.}\ \bibnamefont {McEnally}}, \bibinfo
  {author} {\bibfnamefont {S.}~\bibnamefont {Kim}}, \ and\ \bibinfo {author}
  {\bibfnamefont {L.~D.}\ \bibnamefont {Pfefferle}},\ }\bibfield  {title}
  {\enquote {\bibinfo {title} {Measuring and predicting sooting tendencies of
  oxygenates, alkanes, alkenes, cycloalkanes, and aromatics on a unified
  scale},}\ }\href {\doibase 10.1016/j.combustflame.2017.12.005} {\bibfield
  {journal} {\bibinfo  {journal} {Combust Flame}\ }\textbf {\bibinfo {volume}
  {190}},\ \bibinfo {pages} {349--364} (\bibinfo {year} {2018})}\BibitemShut
  {NoStop}%
\bibitem [{\citenamefont {Janet}\ and\ \citenamefont
  {Kulik}(2017)}]{Janet_2017}%
  \BibitemOpen
  \bibfield  {author} {\bibinfo {author} {\bibfnamefont {J.~P.}\ \bibnamefont
  {Janet}}\ and\ \bibinfo {author} {\bibfnamefont {H.~J.}\ \bibnamefont
  {Kulik}},\ }\bibfield  {title} {\enquote {\bibinfo {title} {Predicting
  electronic structure properties of transition metal complexes with neural
  networks},}\ }\href {\doibase 10.1039/c7sc01247k} {\bibfield  {journal}
  {\bibinfo  {journal} {Chem Sci}\ }\textbf {\bibinfo {volume} {8}},\ \bibinfo
  {pages} {5137–5152} (\bibinfo {year} {2017})}\BibitemShut {NoStop}%
\bibitem [{\citenamefont {Duvenaud}\ \emph {et~al.}(2015)\citenamefont
  {Duvenaud}, \citenamefont {Maclaurin}, \citenamefont {Iparraguirre},
  \citenamefont {Bombarell}, \citenamefont {Hirzel}, \citenamefont
  {Aspuru-Guzik},\ and\ \citenamefont {Adams}}]{Duvenaud2015}%
  \BibitemOpen
  \bibfield  {author} {\bibinfo {author} {\bibfnamefont {D.~K.}\ \bibnamefont
  {Duvenaud}}, \bibinfo {author} {\bibfnamefont {D.}~\bibnamefont {Maclaurin}},
  \bibinfo {author} {\bibfnamefont {J.}~\bibnamefont {Iparraguirre}}, \bibinfo
  {author} {\bibfnamefont {R.}~\bibnamefont {Bombarell}}, \bibinfo {author}
  {\bibfnamefont {T.}~\bibnamefont {Hirzel}}, \bibinfo {author} {\bibfnamefont
  {A.}~\bibnamefont {Aspuru-Guzik}}, \ and\ \bibinfo {author} {\bibfnamefont
  {R.~P.}\ \bibnamefont {Adams}},\ }\bibfield  {title} {\enquote {\bibinfo
  {title} {Convolutional networks on graphs for learning molecular
  fingerprints},}\ }\href@noop {} {\bibfield  {journal} {\bibinfo  {journal}
  {Adv Neural Inf Process Syst}\ }\textbf {\bibinfo {volume} {28}},\ \bibinfo
  {pages} {2224--2232} (\bibinfo {year} {2015})}\BibitemShut {NoStop}%
\bibitem [{\citenamefont {Gilmer}\ \emph {et~al.}(2017)\citenamefont {Gilmer},
  \citenamefont {Schoenholz}, \citenamefont {Riley}, \citenamefont {Vinyals},\
  and\ \citenamefont {Dahl}}]{Gilmer2017}%
  \BibitemOpen
  \bibfield  {author} {\bibinfo {author} {\bibfnamefont {J.}~\bibnamefont
  {Gilmer}}, \bibinfo {author} {\bibfnamefont {S.~S.}\ \bibnamefont
  {Schoenholz}}, \bibinfo {author} {\bibfnamefont {P.~F.}\ \bibnamefont
  {Riley}}, \bibinfo {author} {\bibfnamefont {O.}~\bibnamefont {Vinyals}}, \
  and\ \bibinfo {author} {\bibfnamefont {G.~E.}\ \bibnamefont {Dahl}},\
  }\bibfield  {title} {\enquote {\bibinfo {title} {Neural message passing for
  quantum chemistry},}\ }\href@noop {} {\bibfield  {journal} {\bibinfo
  {journal} {Int Conf Machine Learning}\ } (\bibinfo {year}
  {2017})}\BibitemShut {NoStop}%
\bibitem [{\citenamefont {Battaglia}\ \emph {et~al.}(2018)\citenamefont
  {Battaglia}, \citenamefont {Hamrick}, \citenamefont {Bapst}, \citenamefont
  {Sanchez-Gonzalez}, \citenamefont {Zambaldi}, \citenamefont {Malinowski},
  \citenamefont {Tacchetti}, \citenamefont {Raposo}, \citenamefont {Santoro},
  \citenamefont {Faulkner}, \citenamefont {Gulcehre}, \citenamefont {Song},
  \citenamefont {Ballard}, \citenamefont {Gilmer}, \citenamefont {Dahl},
  \citenamefont {Vaswani}, \citenamefont {Allen}, \citenamefont {Nash},
  \citenamefont {Langston}, \citenamefont {Dyer}, \citenamefont {Heess},
  \citenamefont {Wierstra}, \citenamefont {Kohli}, \citenamefont {Botvinick},
  \citenamefont {Vinyals}, \citenamefont {Li},\ and\ \citenamefont
  {Pascanu}}]{Battaglia_2018}%
  \BibitemOpen
  \bibfield  {author} {\bibinfo {author} {\bibfnamefont {P.~W.}\ \bibnamefont
  {Battaglia}}, \bibinfo {author} {\bibfnamefont {J.~B.}\ \bibnamefont
  {Hamrick}}, \bibinfo {author} {\bibfnamefont {V.}~\bibnamefont {Bapst}},
  \bibinfo {author} {\bibfnamefont {A.}~\bibnamefont {Sanchez-Gonzalez}},
  \bibinfo {author} {\bibfnamefont {V.}~\bibnamefont {Zambaldi}}, \bibinfo
  {author} {\bibfnamefont {M.}~\bibnamefont {Malinowski}}, \bibinfo {author}
  {\bibfnamefont {A.}~\bibnamefont {Tacchetti}}, \bibinfo {author}
  {\bibfnamefont {D.}~\bibnamefont {Raposo}}, \bibinfo {author} {\bibfnamefont
  {A.}~\bibnamefont {Santoro}}, \bibinfo {author} {\bibfnamefont
  {R.}~\bibnamefont {Faulkner}}, \bibinfo {author} {\bibfnamefont
  {C.}~\bibnamefont {Gulcehre}}, \bibinfo {author} {\bibfnamefont
  {F.}~\bibnamefont {Song}}, \bibinfo {author} {\bibfnamefont {A.}~\bibnamefont
  {Ballard}}, \bibinfo {author} {\bibfnamefont {J.}~\bibnamefont {Gilmer}},
  \bibinfo {author} {\bibfnamefont {G.}~\bibnamefont {Dahl}}, \bibinfo {author}
  {\bibfnamefont {A.}~\bibnamefont {Vaswani}}, \bibinfo {author} {\bibfnamefont
  {K.}~\bibnamefont {Allen}}, \bibinfo {author} {\bibfnamefont
  {C.}~\bibnamefont {Nash}}, \bibinfo {author} {\bibfnamefont {V.}~\bibnamefont
  {Langston}}, \bibinfo {author} {\bibfnamefont {C.}~\bibnamefont {Dyer}},
  \bibinfo {author} {\bibfnamefont {N.}~\bibnamefont {Heess}}, \bibinfo
  {author} {\bibfnamefont {D.}~\bibnamefont {Wierstra}}, \bibinfo {author}
  {\bibfnamefont {P.}~\bibnamefont {Kohli}}, \bibinfo {author} {\bibfnamefont
  {M.}~\bibnamefont {Botvinick}}, \bibinfo {author} {\bibfnamefont
  {O.}~\bibnamefont {Vinyals}}, \bibinfo {author} {\bibfnamefont
  {Y.}~\bibnamefont {Li}}, \ and\ \bibinfo {author} {\bibfnamefont
  {R.}~\bibnamefont {Pascanu}},\ }\bibfield  {title} {\enquote {\bibinfo
  {title} {Relational inductive biases, deep learning, and graph networks},}\
  }\href@noop {} {\bibfield  {journal} {\bibinfo  {journal} {arXiv}\ ,\
  \bibinfo {pages} {arXiv:1806.01261}} (\bibinfo {year} {2018})}\BibitemShut
  {NoStop}%
\bibitem [{\citenamefont {Ramakrishnan}\ \emph {et~al.}(2014)\citenamefont
  {Ramakrishnan}, \citenamefont {Dral}, \citenamefont {Rupp},\ and\
  \citenamefont {{von Lilienfeld}}}]{Ramakrishnan_2014}%
  \BibitemOpen
  \bibfield  {author} {\bibinfo {author} {\bibfnamefont {R.}~\bibnamefont
  {Ramakrishnan}}, \bibinfo {author} {\bibfnamefont {P.~O.}\ \bibnamefont
  {Dral}}, \bibinfo {author} {\bibfnamefont {M.}~\bibnamefont {Rupp}}, \ and\
  \bibinfo {author} {\bibfnamefont {O.~A.}\ \bibnamefont {{von Lilienfeld}}},\
  }\bibfield  {title} {\enquote {\bibinfo {title} {Quantum chemistry structures
  and properties of 134 kilo molecules},}\ }\href {\doibase
  10.1038/sdata.2014.22} {\bibfield  {journal} {\bibinfo  {journal} {Sci Data}\
  }\textbf {\bibinfo {volume} {1}} (\bibinfo {year} {2014}),\
  10.1038/sdata.2014.22}\BibitemShut {NoStop}%
\bibitem [{\citenamefont {Sch\"{u}tt}\ \emph
  {et~al.}(2017{\natexlab{a}})\citenamefont {Sch\"{u}tt}, \citenamefont
  {Kindermans}, \citenamefont {Sauceda~Felix}, \citenamefont {Chmiela},
  \citenamefont {Tkatchenko},\ and\ \citenamefont
  {M\"{u}ller}}]{NIPS2017_6700}%
  \BibitemOpen
  \bibfield  {author} {\bibinfo {author} {\bibfnamefont {K.}~\bibnamefont
  {Sch\"{u}tt}}, \bibinfo {author} {\bibfnamefont {P.-J.}\ \bibnamefont
  {Kindermans}}, \bibinfo {author} {\bibfnamefont {H.~E.}\ \bibnamefont
  {Sauceda~Felix}}, \bibinfo {author} {\bibfnamefont {S.}~\bibnamefont
  {Chmiela}}, \bibinfo {author} {\bibfnamefont {A.}~\bibnamefont {Tkatchenko}},
  \ and\ \bibinfo {author} {\bibfnamefont {K.-R.}\ \bibnamefont {M\"{u}ller}},\
  }\bibfield  {title} {\enquote {\bibinfo {title} {Schnet: A continuous-filter
  convolutional neural network for modeling quantum interactions},}\
  }\href@noop {} {\bibfield  {journal} {\bibinfo  {journal} {Adv Neural Inf
  Process Syst}\ }\textbf {\bibinfo {volume} {30}},\ \bibinfo {pages}
  {991--1001} (\bibinfo {year} {2017}{\natexlab{a}})}\BibitemShut {NoStop}%
\bibitem [{\citenamefont {Sch\"{u}tt}\ \emph {et~al.}(2018)\citenamefont
  {Sch\"{u}tt}, \citenamefont {Sauceda}, \citenamefont {Kindermans},
  \citenamefont {Tkatchenko},\ and\ \citenamefont {M\"{u}ller}}]{Sch_tt_2018}%
  \BibitemOpen
  \bibfield  {author} {\bibinfo {author} {\bibfnamefont {K.~T.}\ \bibnamefont
  {Sch\"{u}tt}}, \bibinfo {author} {\bibfnamefont {H.~E.}\ \bibnamefont
  {Sauceda}}, \bibinfo {author} {\bibfnamefont {P.-J.}\ \bibnamefont
  {Kindermans}}, \bibinfo {author} {\bibfnamefont {A.}~\bibnamefont
  {Tkatchenko}}, \ and\ \bibinfo {author} {\bibfnamefont {K.-R.}\ \bibnamefont
  {M\"{u}ller}},\ }\bibfield  {title} {\enquote {\bibinfo {title} {Schnet - a
  deep learning architecture for molecules and materials},}\ }\href {\doibase
  10.1063/1.5019779} {\bibfield  {journal} {\bibinfo  {journal} {J Chem Phys}\
  }\textbf {\bibinfo {volume} {148}},\ \bibinfo {pages} {241722} (\bibinfo
  {year} {2018})}\BibitemShut {NoStop}%
\bibitem [{\citenamefont {J{\o}rgensen}, \citenamefont {Jacobsen},\ and\
  \citenamefont {Schmidt}(2018)}]{Jorgensen_2018}%
  \BibitemOpen
  \bibfield  {author} {\bibinfo {author} {\bibfnamefont {P.~B.}\ \bibnamefont
  {J{\o}rgensen}}, \bibinfo {author} {\bibfnamefont {K.~W.}\ \bibnamefont
  {Jacobsen}}, \ and\ \bibinfo {author} {\bibfnamefont {M.~N.}\ \bibnamefont
  {Schmidt}},\ }\bibfield  {title} {\enquote {\bibinfo {title} {Neural message
  passing with edge updates for predicting properties of molecules and
  materials},}\ }\href@noop {} {\bibfield  {journal} {\bibinfo  {journal}
  {arXiv}\ ,\ \bibinfo {pages} {arXiv:1806.03146}} (\bibinfo {year}
  {2018})}\BibitemShut {NoStop}%
\bibitem [{\citenamefont {LeCun}, \citenamefont {Bengio},\ and\ \citenamefont
  {Hinton}(2015)}]{LeCun_2015}%
  \BibitemOpen
  \bibfield  {author} {\bibinfo {author} {\bibfnamefont {Y.}~\bibnamefont
  {LeCun}}, \bibinfo {author} {\bibfnamefont {Y.}~\bibnamefont {Bengio}}, \
  and\ \bibinfo {author} {\bibfnamefont {G.}~\bibnamefont {Hinton}},\
  }\bibfield  {title} {\enquote {\bibinfo {title} {Deep learning},}\ }\href
  {\doibase 10.1038/nature14539} {\bibfield  {journal} {\bibinfo  {journal}
  {Nature}\ }\textbf {\bibinfo {volume} {521}},\ \bibinfo {pages} {436--444}
  (\bibinfo {year} {2015})}\BibitemShut {NoStop}%
\bibitem [{\citenamefont {Yosinski}\ \emph {et~al.}(2014)\citenamefont
  {Yosinski}, \citenamefont {Clune}, \citenamefont {Bengio},\ and\
  \citenamefont {Lipson}}]{Yosinski2014}%
  \BibitemOpen
  \bibfield  {author} {\bibinfo {author} {\bibfnamefont {J.}~\bibnamefont
  {Yosinski}}, \bibinfo {author} {\bibfnamefont {J.}~\bibnamefont {Clune}},
  \bibinfo {author} {\bibfnamefont {Y.}~\bibnamefont {Bengio}}, \ and\ \bibinfo
  {author} {\bibfnamefont {H.}~\bibnamefont {Lipson}},\ }\bibfield  {title}
  {\enquote {\bibinfo {title} {How transferable are features in deep neural
  networks?}}\ }\href@noop {} {\bibfield  {journal} {\bibinfo  {journal} {Adv
  Neural Inf Process Syst}\ }\textbf {\bibinfo {volume} {27}},\ \bibinfo
  {pages} {3320--3328} (\bibinfo {year} {2014})}\BibitemShut {NoStop}%
\bibitem [{\citenamefont {Smith}, \citenamefont {Isayev},\ and\ \citenamefont
  {Roitberg}(2017)}]{Smith:2017jd}%
  \BibitemOpen
  \bibfield  {author} {\bibinfo {author} {\bibfnamefont {J.~S.}\ \bibnamefont
  {Smith}}, \bibinfo {author} {\bibfnamefont {O.}~\bibnamefont {Isayev}}, \
  and\ \bibinfo {author} {\bibfnamefont {A.~E.}\ \bibnamefont {Roitberg}},\
  }\bibfield  {title} {\enquote {\bibinfo {title} {{ANI}-1: an extensible
  neural network potential with {DFT} accuracy at force field computational
  cost},}\ }\href {\doibase 10.1039/c6sc05720a} {\bibfield  {journal} {\bibinfo
   {journal} {Chem Sci}\ }\textbf {\bibinfo {volume} {8}},\ \bibinfo {pages}
  {3192--3203} (\bibinfo {year} {2017})}\BibitemShut {NoStop}%
\bibitem [{\citenamefont {Yao}\ \emph {et~al.}(2018)\citenamefont {Yao},
  \citenamefont {Herr}, \citenamefont {Toth}, \citenamefont {Mckintyre},\ and\
  \citenamefont {Parkhill}}]{Yao_2018}%
  \BibitemOpen
  \bibfield  {author} {\bibinfo {author} {\bibfnamefont {K.}~\bibnamefont
  {Yao}}, \bibinfo {author} {\bibfnamefont {J.~E.}\ \bibnamefont {Herr}},
  \bibinfo {author} {\bibfnamefont {D.}~\bibnamefont {Toth}}, \bibinfo {author}
  {\bibfnamefont {R.}~\bibnamefont {Mckintyre}}, \ and\ \bibinfo {author}
  {\bibfnamefont {J.}~\bibnamefont {Parkhill}},\ }\bibfield  {title} {\enquote
  {\bibinfo {title} {The tensormol-0.1 model chemistry: a neural network
  augmented with long-range physics},}\ }\href {\doibase 10.1039/c7sc04934j}
  {\bibfield  {journal} {\bibinfo  {journal} {Chem Sci}\ }\textbf {\bibinfo
  {volume} {9}},\ \bibinfo {pages} {2261–2269} (\bibinfo {year}
  {2018})}\BibitemShut {NoStop}%
\bibitem [{\citenamefont {Sch\"{u}tt}\ \emph
  {et~al.}(2017{\natexlab{b}})\citenamefont {Sch\"{u}tt}, \citenamefont
  {Arbabzadah}, \citenamefont {Chmiela}, \citenamefont {M\"{u}ller},\ and\
  \citenamefont {Tkatchenko}}]{Sch_tt_2017}%
  \BibitemOpen
  \bibfield  {author} {\bibinfo {author} {\bibfnamefont {K.~T.}\ \bibnamefont
  {Sch\"{u}tt}}, \bibinfo {author} {\bibfnamefont {F.}~\bibnamefont
  {Arbabzadah}}, \bibinfo {author} {\bibfnamefont {S.}~\bibnamefont {Chmiela}},
  \bibinfo {author} {\bibfnamefont {K.~R.}\ \bibnamefont {M\"{u}ller}}, \ and\
  \bibinfo {author} {\bibfnamefont {A.}~\bibnamefont {Tkatchenko}},\ }\bibfield
   {title} {\enquote {\bibinfo {title} {Quantum-chemical insights from deep
  tensor neural networks},}\ }\href {\doibase 10.1038/ncomms13890} {\bibfield
  {journal} {\bibinfo  {journal} {Nat Commun}\ }\textbf {\bibinfo {volume}
  {8}},\ \bibinfo {pages} {13890} (\bibinfo {year}
  {2017}{\natexlab{b}})}\BibitemShut {NoStop}%
\bibitem [{\citenamefont {Smith}\ \emph {et~al.}(2018)\citenamefont {Smith},
  \citenamefont {Nebgen}, \citenamefont {Zubatyuk}, \citenamefont {Lubbers},
  \citenamefont {Devereux}, \citenamefont {Barros}, \citenamefont {Tretiak},
  \citenamefont {Isayev},\ and\ \citenamefont {Roitberg}}]{Smith_2018}%
  \BibitemOpen
  \bibfield  {author} {\bibinfo {author} {\bibfnamefont {J.~S.}\ \bibnamefont
  {Smith}}, \bibinfo {author} {\bibfnamefont {B.~T.}\ \bibnamefont {Nebgen}},
  \bibinfo {author} {\bibfnamefont {R.}~\bibnamefont {Zubatyuk}}, \bibinfo
  {author} {\bibfnamefont {N.}~\bibnamefont {Lubbers}}, \bibinfo {author}
  {\bibfnamefont {C.}~\bibnamefont {Devereux}}, \bibinfo {author}
  {\bibfnamefont {K.}~\bibnamefont {Barros}}, \bibinfo {author} {\bibfnamefont
  {S.}~\bibnamefont {Tretiak}}, \bibinfo {author} {\bibfnamefont
  {O.}~\bibnamefont {Isayev}}, \ and\ \bibinfo {author} {\bibfnamefont
  {A.}~\bibnamefont {Roitberg}},\ }\bibfield  {title} {\enquote {\bibinfo
  {title} {Outsmarting quantum chemistry through transfer learning},}\
  }\href@noop {} {\bibfield  {journal} {\bibinfo  {journal} {ChemRxiv}\ }
  (\bibinfo {year} {2018})}\BibitemShut {NoStop}%
\bibitem [{\citenamefont {Oosterhout}\ \emph {et~al.}(2015)\citenamefont
  {Oosterhout}, \citenamefont {Kopidakis}, \citenamefont {Owczarczyk},
  \citenamefont {Braunecker}, \citenamefont {Larsen}, \citenamefont
  {Ratcliff},\ and\ \citenamefont {Olson}}]{Oosterhout_2015}%
  \BibitemOpen
  \bibfield  {author} {\bibinfo {author} {\bibfnamefont {S.~D.}\ \bibnamefont
  {Oosterhout}}, \bibinfo {author} {\bibfnamefont {N.}~\bibnamefont
  {Kopidakis}}, \bibinfo {author} {\bibfnamefont {Z.~R.}\ \bibnamefont
  {Owczarczyk}}, \bibinfo {author} {\bibfnamefont {W.~A.}\ \bibnamefont
  {Braunecker}}, \bibinfo {author} {\bibfnamefont {R.~E.}\ \bibnamefont
  {Larsen}}, \bibinfo {author} {\bibfnamefont {E.~L.}\ \bibnamefont
  {Ratcliff}}, \ and\ \bibinfo {author} {\bibfnamefont {D.~C.}\ \bibnamefont
  {Olson}},\ }\bibfield  {title} {\enquote {\bibinfo {title} {Integrating
  theory, synthesis, spectroscopy and device efficiency to design and
  characterize donor materials for organic photovoltaics: a case study
  including 12 donors},}\ }\href {\doibase 10.1039/c5ta01153a} {\bibfield
  {journal} {\bibinfo  {journal} {J Mat Chem A}\ }\textbf {\bibinfo {volume}
  {3}},\ \bibinfo {pages} {9777--9788} (\bibinfo {year} {2015})}\BibitemShut
  {NoStop}%
\bibitem [{\citenamefont {Scharber}\ \emph {et~al.}(2006)\citenamefont
  {Scharber}, \citenamefont {M\"{u}hlbacher}, \citenamefont {Koppe},
  \citenamefont {Denk}, \citenamefont {Waldauf}, \citenamefont {Heeger},\ and\
  \citenamefont {Brabec}}]{Scharber_2006}%
  \BibitemOpen
  \bibfield  {author} {\bibinfo {author} {\bibfnamefont {M.}~\bibnamefont
  {Scharber}}, \bibinfo {author} {\bibfnamefont {D.}~\bibnamefont
  {M\"{u}hlbacher}}, \bibinfo {author} {\bibfnamefont {M.}~\bibnamefont
  {Koppe}}, \bibinfo {author} {\bibfnamefont {P.}~\bibnamefont {Denk}},
  \bibinfo {author} {\bibfnamefont {C.}~\bibnamefont {Waldauf}}, \bibinfo
  {author} {\bibfnamefont {A.}~\bibnamefont {Heeger}}, \ and\ \bibinfo {author}
  {\bibfnamefont {C.}~\bibnamefont {Brabec}},\ }\bibfield  {title} {\enquote
  {\bibinfo {title} {Design rules for donors in bulk-heterojunction solar cells
  - towards 10\% energy-conversion efficiency},}\ }\href {\doibase
  10.1002/adma.200501717} {\bibfield  {journal} {\bibinfo  {journal} {Adv
  Mater}\ }\textbf {\bibinfo {volume} {18}},\ \bibinfo {pages} {789--794}
  (\bibinfo {year} {2006})}\BibitemShut {NoStop}%
\bibitem [{\citenamefont {Li}, \citenamefont {McCulloch},\ and\ \citenamefont
  {Brabec}(2018)}]{Li_2018}%
  \BibitemOpen
  \bibfield  {author} {\bibinfo {author} {\bibfnamefont {N.}~\bibnamefont
  {Li}}, \bibinfo {author} {\bibfnamefont {I.}~\bibnamefont {McCulloch}}, \
  and\ \bibinfo {author} {\bibfnamefont {C.~J.}\ \bibnamefont {Brabec}},\
  }\bibfield  {title} {\enquote {\bibinfo {title} {Analyzing the efficiency,
  stability and cost potential for fullerene-free organic photovoltaics in one
  figure of merit},}\ }\href {\doibase 10.1039/c8ee00151k} {\bibfield
  {journal} {\bibinfo  {journal} {Energy Environ Sci}\ }\textbf {\bibinfo
  {volume} {11}},\ \bibinfo {pages} {1355--1361} (\bibinfo {year}
  {2018})}\BibitemShut {NoStop}%
\bibitem [{\citenamefont {Kanal}\ \emph {et~al.}(2013)\citenamefont {Kanal},
  \citenamefont {Owens}, \citenamefont {Bechtel},\ and\ \citenamefont
  {Hutchison}}]{Kanal_2013}%
  \BibitemOpen
  \bibfield  {author} {\bibinfo {author} {\bibfnamefont {I.~Y.}\ \bibnamefont
  {Kanal}}, \bibinfo {author} {\bibfnamefont {S.~G.}\ \bibnamefont {Owens}},
  \bibinfo {author} {\bibfnamefont {J.~S.}\ \bibnamefont {Bechtel}}, \ and\
  \bibinfo {author} {\bibfnamefont {G.~R.}\ \bibnamefont {Hutchison}},\
  }\bibfield  {title} {\enquote {\bibinfo {title} {Efficient computational
  screening of organic polymer photovoltaics},}\ }\href {\doibase
  10.1021/jz400215j} {\bibfield  {journal} {\bibinfo  {journal} {The Journal of
  Physical Chemistry Letters}\ }\textbf {\bibinfo {volume} {4}},\ \bibinfo
  {pages} {1613–1623} (\bibinfo {year} {2013})}\BibitemShut {NoStop}%
\bibitem [{\citenamefont {Pyzer-Knapp}, \citenamefont {Li},\ and\ \citenamefont
  {Aspuru-Guzik}(2015)}]{Pyzer_Knapp_2015}%
  \BibitemOpen
  \bibfield  {author} {\bibinfo {author} {\bibfnamefont {E.~O.}\ \bibnamefont
  {Pyzer-Knapp}}, \bibinfo {author} {\bibfnamefont {K.}~\bibnamefont {Li}}, \
  and\ \bibinfo {author} {\bibfnamefont {A.}~\bibnamefont {Aspuru-Guzik}},\
  }\bibfield  {title} {\enquote {\bibinfo {title} {Learning from the harvard
  clean energy project: The use of neural networks to accelerate materials
  discovery},}\ }\href {\doibase 10.1002/adfm.201501919} {\bibfield  {journal}
  {\bibinfo  {journal} {Advanced Functional Materials}\ }\textbf {\bibinfo
  {volume} {25}},\ \bibinfo {pages} {6495–6502} (\bibinfo {year}
  {2015})}\BibitemShut {NoStop}%
\bibitem [{\citenamefont {Jørgensen}\ \emph {et~al.}(2018)\citenamefont
  {Jørgensen}, \citenamefont {Mesta}, \citenamefont {Shil}, \citenamefont
  {García~Lastra}, \citenamefont {Jacobsen}, \citenamefont {Thygesen},\ and\
  \citenamefont {Schmidt}}]{J_rgensen_2018}%
  \BibitemOpen
  \bibfield  {author} {\bibinfo {author} {\bibfnamefont {P.~B.}\ \bibnamefont
  {Jørgensen}}, \bibinfo {author} {\bibfnamefont {M.}~\bibnamefont {Mesta}},
  \bibinfo {author} {\bibfnamefont {S.}~\bibnamefont {Shil}}, \bibinfo {author}
  {\bibfnamefont {J.~M.}\ \bibnamefont {García~Lastra}}, \bibinfo {author}
  {\bibfnamefont {K.~W.}\ \bibnamefont {Jacobsen}}, \bibinfo {author}
  {\bibfnamefont {K.~S.}\ \bibnamefont {Thygesen}}, \ and\ \bibinfo {author}
  {\bibfnamefont {M.~N.}\ \bibnamefont {Schmidt}},\ }\bibfield  {title}
  {\enquote {\bibinfo {title} {Machine learning-based screening of complex
  molecules for polymer solar cells},}\ }\href {\doibase 10.1063/1.5023563}
  {\bibfield  {journal} {\bibinfo  {journal} {The Journal of Chemical Physics}\
  }\textbf {\bibinfo {volume} {148}},\ \bibinfo {pages} {241735} (\bibinfo
  {year} {2018})}\BibitemShut {NoStop}%
\bibitem [{\citenamefont {Frisch}\ \emph {et~al.}()\citenamefont {Frisch},
  \citenamefont {Trucks}, \citenamefont {Schlegel}, \citenamefont {Scuseria},
  \citenamefont {Robb}, \citenamefont {Cheeseman}, \citenamefont {Scalmani},
  \citenamefont {Barone}, \citenamefont {Mennucci} \emph
  {et~al.}}]{Gaussian2009}%
  \BibitemOpen
  \bibfield  {author} {\bibinfo {author} {\bibfnamefont {M.~J.}\ \bibnamefont
  {Frisch}}, \bibinfo {author} {\bibfnamefont {G.~W.}\ \bibnamefont {Trucks}},
  \bibinfo {author} {\bibfnamefont {H.~B.}\ \bibnamefont {Schlegel}}, \bibinfo
  {author} {\bibfnamefont {G.~E.}\ \bibnamefont {Scuseria}}, \bibinfo {author}
  {\bibfnamefont {M.~A.}\ \bibnamefont {Robb}}, \bibinfo {author}
  {\bibfnamefont {J.~R.}\ \bibnamefont {Cheeseman}}, \bibinfo {author}
  {\bibfnamefont {G.}~\bibnamefont {Scalmani}}, \bibinfo {author}
  {\bibfnamefont {V.}~\bibnamefont {Barone}}, \bibinfo {author} {\bibfnamefont
  {B.}~\bibnamefont {Mennucci}},  \emph {et~al.},\ }\href@noop {} {\enquote
  {\bibinfo {title} {Gaussian~09 {R}evision {D}.1},}\ }\bibinfo {note}
  {Gaussian Inc. Wallingford CT 2009}\BibitemShut {NoStop}%
\bibitem [{\citenamefont {Larsen}(2016)}]{Larsen_2016}%
  \BibitemOpen
  \bibfield  {author} {\bibinfo {author} {\bibfnamefont {R.~E.}\ \bibnamefont
  {Larsen}},\ }\bibfield  {title} {\enquote {\bibinfo {title} {Simple
  extrapolation method to predict the electronic structure of conjugated
  polymers from calculations on oligomers},}\ }\href {\doibase
  10.1021/acs.jpcc.6b02138} {\bibfield  {journal} {\bibinfo  {journal} {J Phys
  Chem C}\ }\textbf {\bibinfo {volume} {120}},\ \bibinfo {pages} {9650--9660}
  (\bibinfo {year} {2016})}\BibitemShut {NoStop}%
\bibitem [{\citenamefont {Weininger}(1988)}]{Weininger_1988}%
  \BibitemOpen
  \bibfield  {author} {\bibinfo {author} {\bibfnamefont {D.}~\bibnamefont
  {Weininger}},\ }\bibfield  {title} {\enquote {\bibinfo {title} {Smiles, a
  chemical language and information system. 1. introduction to methodology and
  encoding rules},}\ }\href {\doibase 10.1021/ci00057a005} {\bibfield
  {journal} {\bibinfo  {journal} {J Chem Inf Model}\ }\textbf {\bibinfo
  {volume} {28}},\ \bibinfo {pages} {31--36} (\bibinfo {year}
  {1988})}\BibitemShut {NoStop}%
\bibitem [{\citenamefont {Lloyd}, \citenamefont {Anthony},\ and\ \citenamefont
  {Malliaras}(2007)}]{Lloyd_2007}%
  \BibitemOpen
  \bibfield  {author} {\bibinfo {author} {\bibfnamefont {M.~T.}\ \bibnamefont
  {Lloyd}}, \bibinfo {author} {\bibfnamefont {J.~E.}\ \bibnamefont {Anthony}},
  \ and\ \bibinfo {author} {\bibfnamefont {G.~G.}\ \bibnamefont {Malliaras}},\
  }\bibfield  {title} {\enquote {\bibinfo {title} {Photovoltaics from soluble
  small molecules},}\ }\href {\doibase 10.1016/s1369-7021(07)70277-8}
  {\bibfield  {journal} {\bibinfo  {journal} {Mater Today}\ }\textbf {\bibinfo
  {volume} {10}},\ \bibinfo {pages} {34--41} (\bibinfo {year}
  {2007})}\BibitemShut {NoStop}%
\bibitem [{\citenamefont {van~der Poll}\ \emph {et~al.}(2012)\citenamefont
  {van~der Poll}, \citenamefont {Love}, \citenamefont {Nguyen},\ and\
  \citenamefont {Bazan}}]{van_der_Poll_2012}%
  \BibitemOpen
  \bibfield  {author} {\bibinfo {author} {\bibfnamefont {T.~S.}\ \bibnamefont
  {van~der Poll}}, \bibinfo {author} {\bibfnamefont {J.~A.}\ \bibnamefont
  {Love}}, \bibinfo {author} {\bibfnamefont {T.-Q.}\ \bibnamefont {Nguyen}}, \
  and\ \bibinfo {author} {\bibfnamefont {G.~C.}\ \bibnamefont {Bazan}},\
  }\bibfield  {title} {\enquote {\bibinfo {title} {Non-basic high-performance
  molecules for solution-processed organic solar cells},}\ }\href {\doibase
  10.1002/adma.201201127} {\bibfield  {journal} {\bibinfo  {journal} {Adv
  Mater}\ }\textbf {\bibinfo {volume} {24}},\ \bibinfo {pages} {3646--3649}
  (\bibinfo {year} {2012})}\BibitemShut {NoStop}%
\bibitem [{\citenamefont {{Li}}\ \emph {et~al.}(2016)\citenamefont {{Li}},
  \citenamefont {{Tarlow}}, \citenamefont {{Brockschmidt}},\ and\ \citenamefont
  {{Zemel}}}]{Li2016}%
  \BibitemOpen
  \bibfield  {author} {\bibinfo {author} {\bibfnamefont {Y.}~\bibnamefont
  {{Li}}}, \bibinfo {author} {\bibfnamefont {D.}~\bibnamefont {{Tarlow}}},
  \bibinfo {author} {\bibfnamefont {M.}~\bibnamefont {{Brockschmidt}}}, \ and\
  \bibinfo {author} {\bibfnamefont {R.}~\bibnamefont {{Zemel}}},\ }\bibfield
  {title} {\enquote {\bibinfo {title} {Gated graph sequence neural networks},}\
  }\href@noop {} {\bibfield  {journal} {\bibinfo  {journal} {Int Conf Learn
  Represent}\ } (\bibinfo {year} {2016})}\BibitemShut {NoStop}%
\bibitem [{\citenamefont {Rappe}\ \emph {et~al.}(1992)\citenamefont {Rappe},
  \citenamefont {Casewit}, \citenamefont {Colwell}, \citenamefont {Goddard},\
  and\ \citenamefont {Skiff}}]{Rappe_1992}%
  \BibitemOpen
  \bibfield  {author} {\bibinfo {author} {\bibfnamefont {A.~K.}\ \bibnamefont
  {Rappe}}, \bibinfo {author} {\bibfnamefont {C.~J.}\ \bibnamefont {Casewit}},
  \bibinfo {author} {\bibfnamefont {K.~S.}\ \bibnamefont {Colwell}}, \bibinfo
  {author} {\bibfnamefont {W.~A.}\ \bibnamefont {Goddard}}, \ and\ \bibinfo
  {author} {\bibfnamefont {W.~M.}\ \bibnamefont {Skiff}},\ }\bibfield  {title}
  {\enquote {\bibinfo {title} {Uff, a full periodic table force field for
  molecular mechanics and molecular dynamics simulations},}\ }\href {\doibase
  10.1021/ja00051a040} {\bibfield  {journal} {\bibinfo  {journal} {Journal of
  the American Chemical Society}\ }\textbf {\bibinfo {volume} {114}},\ \bibinfo
  {pages} {10024–10035} (\bibinfo {year} {1992})}\BibitemShut {NoStop}%
\bibitem [{\citenamefont {Larsen}\ \emph {et~al.}()\citenamefont {Larsen},
  \citenamefont {Olson}, \citenamefont {Kopidakis}, \citenamefont {Owczarczyk},
  \citenamefont {Hammond}, \citenamefont {Graf}, \citenamefont {Kemper},
  \citenamefont {Sides}, \citenamefont {Munch}, \citenamefont {Evenson},\ and\
  \citenamefont {Swank}}]{opv}%
  \BibitemOpen
  \bibfield  {author} {\bibinfo {author} {\bibfnamefont {R.}~\bibnamefont
  {Larsen}}, \bibinfo {author} {\bibfnamefont {D.}~\bibnamefont {Olson}},
  \bibinfo {author} {\bibfnamefont {N.}~\bibnamefont {Kopidakis}}, \bibinfo
  {author} {\bibfnamefont {Z.}~\bibnamefont {Owczarczyk}}, \bibinfo {author}
  {\bibfnamefont {S.}~\bibnamefont {Hammond}}, \bibinfo {author} {\bibfnamefont
  {P.}~\bibnamefont {Graf}}, \bibinfo {author} {\bibfnamefont {T.}~\bibnamefont
  {Kemper}}, \bibinfo {author} {\bibfnamefont {S.}~\bibnamefont {Sides}},
  \bibinfo {author} {\bibfnamefont {K.}~\bibnamefont {Munch}}, \bibinfo
  {author} {\bibfnamefont {D.}~\bibnamefont {Evenson}}, \ and\ \bibinfo
  {author} {\bibfnamefont {C.}~\bibnamefont {Swank}},\ }\href@noop {} {\enquote
  {\bibinfo {title} {Computational database for active layer materials for
  organic photovoltaic solar cells},}\ }\BibitemShut {NoStop}%
\end{thebibliography}%

\end{document}